# Conducting interface in oxide homojunction: understanding of superior properties in black TiO$_2$

*Xujie Lü,\*,†,‡,∥ Aiping Chen,†,∥ Yongkang Luo,† Ping Lu,§ Yaomin Dai,† Erik Enriquez,† Paul Dowden,† Hongwu Xu,\*,‡ Paul G. Kotula,§ Abul K. Azad,† Dmitry A. Yarotski,† Rohit P. Prasankumar,† Antoinette J. Taylor,† Joe D. Thompson,† and Quanxi Jia\*,†*

†Materials Physics and Applications Division, Los Alamos National Laboratory, Los Alamos, New Mexico 87545, United States

‡Earth and Environmental Sciences Division, Los Alamos National Laboratory, Los Alamos, New Mexico 87545, United States

§Sandia National Laboratories, Albuquerque, New Mexico 87185, United States

ABSTRACT. Black TiO$_2$ nanoparticles with a crystalline-core and amorphous-shell structure exhibit superior optoelectronic properties in comparison with pristine TiO$_2$. The fundamental mechanisms underlying these enhancements, however, remain unclear, largely due to the inherent complexities and limitations of powder materials. Here, we fabricate TiO$_2$ homojunction films consisting of an oxygen-deficient amorphous layer on top of a highly crystalline layer, to simulate the structural/functional configuration of black TiO$_2$ nanoparticles. Metallic conduction is achieved at the crystalline-amorphous homointerface via electronic interface reconstruction, which we show to be the main reason for the superior properties of black TiO$_2$. This work not only achieves an unprecedented understanding of black TiO$_2$, but also provides a new perspective for investigating carrier generation and transport behavior at oxide interfaces, which are of tremendous fundamental and technological interest.

As one of the most attractive functional materials, titanium dioxide ($TiO_2$) has been investigated extensively in advanced energy conversion and storage fields,[1-5] where $TiO_2$ is required to possess outstanding functionalities of light absorption, charge separation and electron transport. The wide band gap (> 3.0 eV) and relatively low conductivity, however, limit its technological applications. Numerous efforts have been made to improve these properties, including ion-doping, semiconductor-compositing, and nanofabrication.[6-8] Recently, an appealing approach based on disorder engineering was proposed, resulting in a new form of $TiO_2$ nanoparticles with black color[9]. Such black $TiO_2$ exceeds the limitations of pristine $TiO_2$ by showing much improved performance in important applications, such as photocatalytic hydrogen generation/water decontamination,[9] rechargeable batteries,[10] and supercapacitors.[11] These improved properties have been demonstrated to largely benefit from the enhanced electron transport of black $TiO_2$.[12-14] Therefore, understanding the origin of enhanced electron transport would hold the key to the mechanism of the superior properties.

Black $TiO_2$ nanomaterials (*e.g.* nanoparticles, nanotubes) have been synthesized by various methods, including high-pressure hydrogenation, hydrogen-plasma assisted reaction and aluminum reduction.[9,15,16] Although these materials have demonstrated improved performance, the explanations are largely diverse.[9,15-20] Some studies attribute to the existence of hydrogen,[9,15,19] while others reported that the samples prepared by non-hydrogen methods also exhibit substantially enhanced properties.[16,18,20] Interestingly, the black $TiO_2$ materials prepared by different synthetic methods share a unique core-shell structure with an oxygen-deficient disordered shell and a crystalline core, which is believed to be a key factor.[12,17,18] However, owing to the complicated effects from a number of factors in powder materials, such as size distributions, grain boundaries and hydrogen incorporation, the role of the core-shell structure in the improved properties of black $TiO_2$ remain unclear.

Here, we fabricate the black $TiO_2$ structure into a thin-film form in a controlled manner, which can exclude the aforementioned influencing factors and thus brings a simpler system with excellent reproducibility, to facilitate understanding the origin of the enhanced properties. Specifically, $TiO_2$ bilayer thin films with a crystalline layer (core) covered by an oxygen-deficient amorphous layer (shell) have been designed to simulate the structural/functional configuration of black $TiO_2$ using pulsed laser deposition (PLD), as illustrated in Figure 1a.

Strikingly, metallic conduction is achieved at the crystalline-amorphous interface in the bilayer films, which is responsible for the enhanced electron transport of black TiO$_2$. The mechanism of the interfacial metallic conduction has been revealed from systematic experiments, including Hall measurements, terahertz (THz) spectroscopy, scanning transmission electron microscopy (STEM) and electron energy-loss spectroscopy (EELS), that demonstrate the dominant role of an electronically reconstructed homointerface.

Bilayer TiO$_2$ films were constructed by depositing a crystalline layer at 600°C, followed by growing an amorphous layer at 100°C. Single-layer crystalline and amorphous films were also deposited for comparison. Experimental details can be found in the Supporting Information. Figure 1b shows the X-ray diffraction (XRD) patterns of these TiO$_2$ films on LaAlO$_3$ (LAO) substrates. The anatase TiO$_2$ (004) peak appears in both bilayer and single-layer crystalline films, indicating their oriented growth. By contrast, the single-layer film grown at 100°C does not show any TiO$_2$ diffraction peak, indicating its amorphous nature. In-situ reflection high-energy electron diffraction (RHEED) reveals the high crystallinity of the crystalline layer (Figure 1c). X-ray reflectivity (XRR) measurements show that the roughness of crystalline-amorphous interface is about 1~2 unit cells (Figure S1). STEM was used to characterize the microstructure and interface of the bilayer film. Cross-sectional high-angle annular dark field (HAADF) STEM images, shown in Figures 1d&e, confirm the high quality of the crystalline layer and the disordered feature of the amorphous layer, whose thicknesses are 22 and 15 nm, respectively.

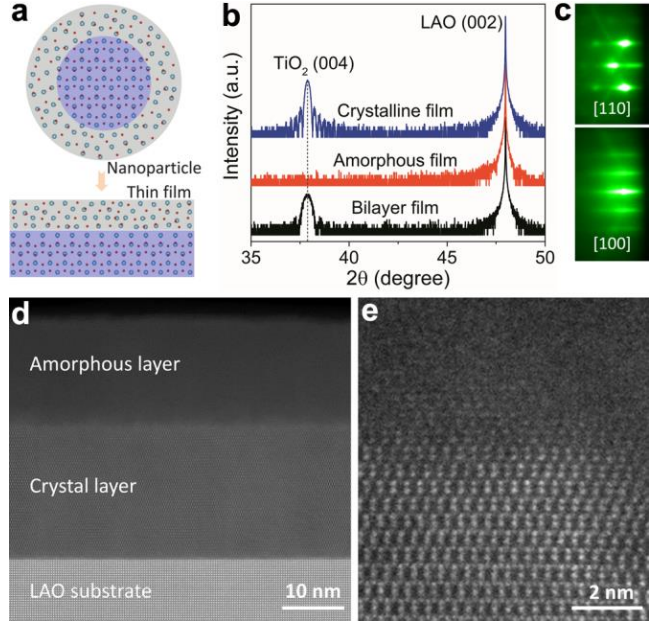

**Figure 1.** (a) Illustration of black $TiO_2$ nanoparticle and thin film with crystalline-amorphous bilayer structure. (b) XRD 2θ-ω scans of $TiO_2$ films on LAO (001). (c) RHEED patterns taken during growth of the $TiO_2$ crystalline layer. The roughness of crystalline-amorphous interface is determined to be 1~2 unit cells by X-ray reflectivity (Figure S1). (d) Cross-sectional HAADF-STEM image of the bilayer film. (e) Magnified STEM image at the interface.

Figure 2 presents the temperature-dependent electrical resistivity ($\rho_{xx}$) and Hall resistivity ($\rho_{yx}$) of the $TiO_2$ films. As is seen, the three types of films show dramatically different transport properties. At room temperature (RT), the resistivity of the bilayer $TiO_2$ film is $5.9 \times 10^{-3}$ Ω cm, which is more than five (two) orders of magnitude smaller than that of the crystalline (amorphous) film. More importantly, the resistivity of bilayer $TiO_2$ decreases upon cooling from RT, a characteristic of metallic behavior. In contrast, both the crystalline and amorphous films exhibit insulating behavior, with $\rho_{xx}(T)$ increasing exponentially with decreasing temperature. This implies a metallic conducting interface between the crystalline and amorphous layers, which will be systemically elaborated on later. Note that $\rho_{xx}(T)$ of bilayer $TiO_2$ increases when $T < 110$ K. Such a behavior cannot be interpreted by the Arrhenius relation [$\rho_{xx} \propto \exp(E_a/k_BT)$, where $E_a$ is the activation energy and $k_B$ is Boltzmann constant], but well conforms to a $-\log T$ dependence (cf. the inset of Figure 2a), suggesting that the increase in resistivity at low temperatures is probably caused by electron localization due to disorder potentials.[21]

In the bilayer film, two possible reasons for its high conductivity could be (i) oxygen vacancies in the amorphous layer, and (ii) electron transport at the crystalline-TiO$_2$/LAO interface. The former possibility is ruled out by comparing its temperature-dependent resistivities with that of the single-layer amorphous TiO$_2$ film (Figure 2a), which has two orders of magnitude higher resistivity at RT and shows a typical insulating behavior. The latter one can easily be eliminated due to the high resistivity (410 Ω cm at RT) and insulating $R$–$T$ behavior of the single-layer crystalline film. A simple parallel-circuit contribution also is unlikely, for which we calculated the shunt resistivity of the bilayer film via $R_S = (R_c \times R_a)/(R_c+R_a)$, where $R_S$, $R_c$ and $R_a$ are the shunt, crystalline- and amorphous-layer resistances, respectively. The calculated shunt resistivities, shown as open circles in Figure 2a, are remarkably different from the measured values, in both magnitude and temperature dependency. Therefore, we infer that metallic conduction in the bilayer TiO$_2$ film originates from the formation of a conducting channel at the interface between the crystalline and amorphous layers. To prove this point, we introduce a conducting interface channel to the parallel model, and the fitting curve (orange line in Figure 2a) accurately reproduces $\rho_{xx}(T)$ of the bilayer film, as discussed in detail in the Supporting Information (Figure S2). Further, a TiO$_2$ bilayer film with doubled thicknesses for both layers shows similar sheet resistances (Figure S3), confirming its interface-dominated conduction.

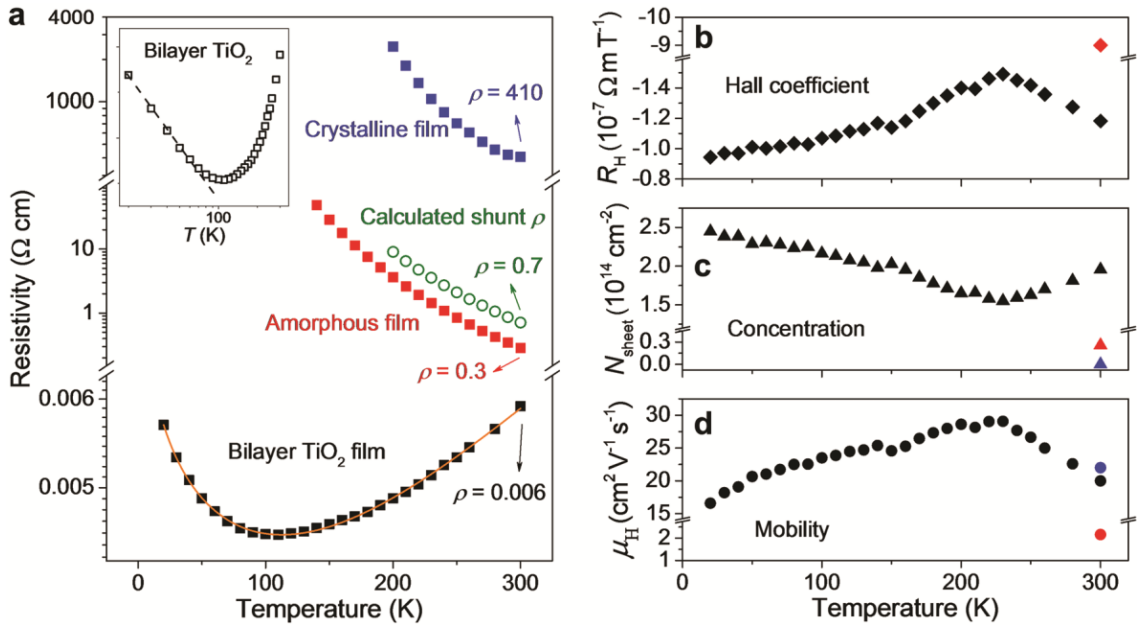

**Figure 2.** (a) Temperature-dependent resistivity of the bilayer $TiO_2$ film (black) in comparison with the single-layer crystalline (blue) and amorphous (red) films. Open circles show the calculated shunt resistivities of the bilayer film without considering the interface contribution. The orange line is the fitted $\rho_{xx}(T)$ curve using a parallel conduction channel model. The inset displays the resistivity of the bilayer film plotted on a $\log T$ scale. (b–d) Temperature-dependent Hall coefficient ($R_H$), sheet electron concentration ($N_{sheet}$) and Hall mobility ($\mu_H$) of the bilayer film. The blue and red symbols indicate corresponding values for the single-layer crystalline and amorphous films, respectively.

The Hall resistivity is negative and linear with magnetic field, manifesting an electron-dominant conduction. Figures 2B−2D present the temperature-dependent Hall coefficient $R_H = \rho_{yx}/B$, sheet carrier concentration ($N_{sheet}$) and Hall mobility ($\mu_H$) of the bilayer $TiO_2$ film. At room temperature, $N_{sheet}$ and $\mu_H$ are $\sim 10^{14}$ cm$^{-2}$ and 20 cm$^2$V$^{-1}$s$^{-1}$, respectively. The carrier concentration of bilayer $TiO_2$ is much higher than that of both the crystalline ($\sim 10^9$ cm$^{-2}$) and amorphous ($\sim 10^{13}$ cm$^{-2}$) films. Its mobility at RT is ten times higher than that of the amorphous film ($\sim 2$ cm$^2$V$^{-1}$s$^{-1}$), but is comparable to that of the crystalline film (22 cm$^2$V$^{-1}$s$^{-1}$). Generally, $N_{sheet}$ decreases with increasing temperature, contrary to an insulator (or semiconductor) in which the carrier concentration increases exponentially as temperature rises.[21] This further demonstrates the metallic conduction of the bilayer $TiO_2$ film. Note that the mobility of the bilayer film first increases with decreasing temperature, reaches a maximum value of $\sim 30$ cm$^2$V$^{-1}$s$^{-1}$, and decreases thereafter. Such a $\mu_H(T)$ profile is not surprising considering the aforementioned disorder-induced electron localization. Significantly, by constructing such a homointerface, the electron concentration and mobility in the bilayer $TiO_2$ films are comparable to values obtained in $TiO_2$-based conducting oxides via transition-metal doping (Table S2). These results suggest that high carrier concentration and mobility can be achieved concurrently at oxide homointerfaces.

Optical conductivity, with its zero-frequency extrapolation being the direct-current conductivity, is sensitive to the presence of metallic behavior, which usually gives rise to a Drude response in the low-frequency spectrum.[22] To further confirm metallicity of the bilayer film, we measured the optical conductivity in the THz range. As shown in Figure 3, a Drude response is observed in the THz spectrum of the bilayer film. We fitted the experimental data to the Drude model,

$$\sigma_1(\omega) = \frac{2\pi}{Z_0} \frac{\Omega_p^2}{\tau(\omega^2 + \tau^{-2})}$$

where $Z_0$ (≈ 377Ω) is the vacuum impedance, $\Omega_p$ and $1/\tau$ correspond to the plasma frequency and scattering rate of the free carriers, respectively. The low-frequency Drude response in the bilayer film clearly signifies the presence of metallic state. Note that the zero-frequency value of the Drude fit agrees well with the transport measurements (open square). In contrast, the optical conductivities of the amorphous and crystalline films are vanishingly small and basically frequency independent. The absence of any Drude-like behavior signals their non-metallic nature. Furthermore, the optical conductivities were measured at different temperatures for the bilayer film, where a Drude response appears at each temperature down to 25 K (Figure S4). The temperature-dependent resistivities derived from the zero-frequency values of the Drude profiles are shown as red squares in the inset of Figure 3, consistent with the transport measurements (black line).

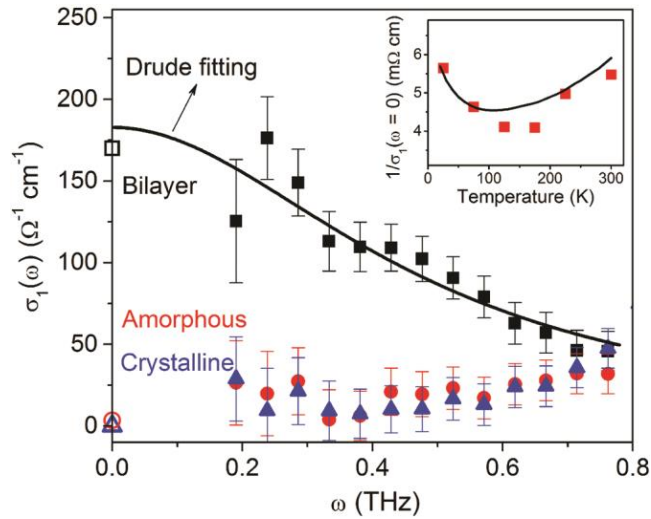

**Figure 3.** Terahertz optical conductivities of TiO$_2$ bilayer (black), crystalline (blue) and amorphous (red) films at RT. The solid line is a Drude fitting to data for the bilayer film. The open symbols at zero frequency represent the conductivities derived from transport

measurements. The inset compares the temperature-dependent resistivities obtained by THz (red squares) and transport (black line) measurements.

These intriguing findings drove us to further investigate the origin of the interface-induced metallic behavior in the bilayer $TiO_2$ film. The conducting crystalline-amorphous homointerface is responsible for the enhanced electron conduction of black $TiO_2$ with crystalline-core and amorphous-shell structure. Emergent properties have been generated at interfaces between oxides,[23-27] such as interfacial superconductivity[25] and a high-mobility electron gas at LAO/SrTiO$_3$ (STO) interface.[23] However, interface-induced emergent properties have only been investigated well in heterostructures, and highly depend on the surface termination.[23] For instance, in the LAO/STO system, $TiO_2$-terminated STO induces a conducting interface while SrO termination brings an insulating interface. Our work thus provides a simpler system, a $TiO_2$ homostructure, for investigating carrier generation and transport at oxide interfaces, and demonstrate that oxide homointerfaces can also induce emergent phenomena.

Having demonstrated interface-induced metallic conduction in the bilayer $TiO_2$ film, we trace the underlying mechanisms. In oxides, compositional compensation is not the only option for charge rearrangement: mixed-valence charge compensation occurs if electrons can be redistributed at a lower energy cost than that for redistributing ions.[28] In the bilayer $TiO_2$ system, oxygen vacancies exist in the amorphous layer because of the oxygen-deficient growth condition ($10^{-6}$ Torr); meanwhile, Ti has accessible mixed-valence character, allowing for reduction from $Ti^{4+}$ towards $Ti^{3+}$. Thus, the amorphous layer is denoted as oxygen-deficient $Ti^{(4-2x)+}O_{2-x}$. For the crystalline layer, its high crystallinity ensures low defect content, thus low carrier density but relatively high electron mobility (Figure 2). We have probed the microscopic distribution of formal valences using EELS in the STEM, where Ti−$L$ and O−$K$ edges were recorded simultaneously (Figure 4a&b, and Figure S5). The Ti−$L$ edge provides a fingerprint of the $Ti^{4+}$ and $Ti^{3+}$ states, and the O−$K$ edge is sensitive to the presence of oxygen vacancies.[28,29] EELS spectra collected close to (0.75, 1.5, 2.25 and 3.0 nm) and far away (7.5 nm) from the interface are displayed in Figure 4b, which shows that the Ti−$L_2$ peak shifts by 0.5 eV from the crystalline side to amorphous side. The change of the Ti−$L$ energy demonstrates the existence of $Ti^{3+}$ in the amorphous layer, at the interface, and in the crystalline side close to the interface (within 3 nm). In contrast, the O−$K$ EELS spectrum collected in the crystalline side close to the interface (0.75 nm) is almost the same as that collected far away from the interface, suggesting that the interface at the crystalline layer side possesses few oxygen vacancies, as expected for the high-crystalline TiO2.

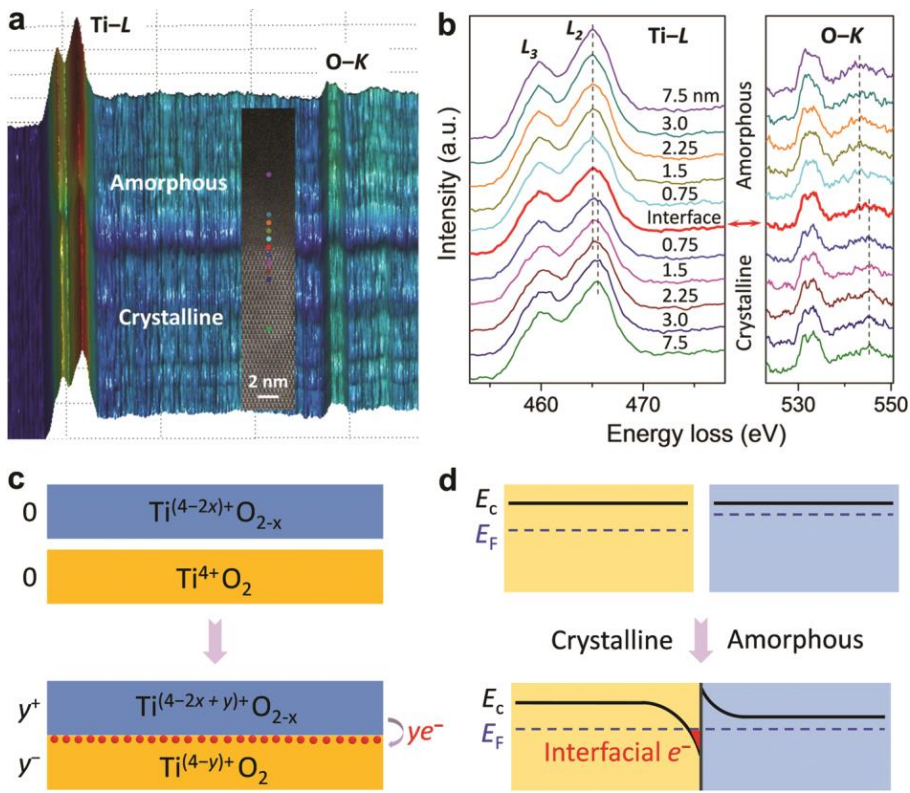

**Figure 4.** Interfacial properties of the TiO2 bilayer film. (a) 3D EELS spectra displaced relative to the specimen position. Inset shows the corresponding STEM image. (b) Selected Ti−$L$ and O−$K$ EELS spectra recorded at various positions as marked by colored dots in the STEM image of panel a. Schematic illustrations of (c) electronic interfacial reconstruction and (d) band diagram of the TiO2 homointerface before and after reconstruction.

Conceptually, one can construct the interface from two neutral units and then allow charge transfer between them. Defects in the amorphous layer (*e.g.* disorder and oxygen vacancies) form a narrow band of states close to the bottom of the conduction band (CB) where the Fermi level ($E_F$) is pinned.[21] $E_F$ in the amorphous layer is closer to CB than it is in the crystalline layer. When they contact, band bending occurs to balance their Fermi levels, which allows transfer of electrons from the amorphous layer to the crystalline layer, as illustrated in Figure 4c&d. This process renders the amorphous side positive and crystalline side negative, keeping an overall neutral charge. Such a situation is similar to a high-low junction (or $n-n^+$ homojunction), where the transferred electrons are confined around the interface.[30,31] In other words, electrons accumulate at the interface on the crystalline side within ~3 nm (Figure 4b), resulting in a high interfacial carrier concentration. These interfacial charges are compensated by the reduced $Ti^{3+}$ that places extra electrons in the $TiO_2$ CB to occupy Ti 3$d$ states, undergoing an electronic interfacial reconstruction.[32] Experimentally, these interfacial electrons also possess a high mobility as in crystalline $TiO_2$, which has been proven by transport measurements (Figure 2 and Table S2). In this case, the two functions of carrier supply and transport are spatially separated in the bilayer film: the amorphous layer provides electrons and the interfacial area of the crystalline layer acts as the electron transport channel. This design enables both high carrier concentration and high mobility, which are concurrently achieved at the homointerface to gain superior collective properties.

In summary, the findings presented here not only provide an unprecedented understanding of black $TiO_2$ and the mechanism of its enhanced electron transport, but also open a new route for developing novel highly conducting oxides and their films by constructing a homointerface with mixed valences. By engineering the interface, high electron concentration and mobility have been concurrently achieved at the homointerface between crystalline and amorphous layers in a bilayer $TiO_2$ thin film. The RT resistivity of the bilayer film is measured to be $5.9 \times 10^{-3}$ Ω cm, which is several orders of magnitude lower than that of the single-layer crystalline and amorphous films. The remarkable electronic properties generated at the $TiO_2$ homointerface provide a simpler and cleaner system for investigating the carrier generation and transport behavior at oxide interfaces. The interfacial engineered $TiO_2$ thin films possessing metallic conduction are promising for various applications such as thin-film transistors, low-cost LEDs and high-performance photovoltaic devices. The exploration of unique interfaces in oxides is still in its infancy; more extraordinary phenomena, such as high-mobility 2D electron gases and interfacial superconductivity, could be generated in homointerfaces by selecting suitable systems, optimizing growth conditions, or making a gate electrode in electronic devices.

ASSOCIATED CONTENT

**Supporting Information**. Experimental details of preparation and characterization for TiO$_2$ films, comparison of properties of various films, X-ray reflectivity pattern of the bilayer film, sheet resistance of TiO$_2$ bilayer films film different thicknesses, THz optical conductivities of the bilayer film at various temperatures, 46 EELS spectra recorded for the bilayer film, electrical field transmitted through the TiO$_2$ bilayer film as a function of gate delay, and detailed discussion of the $\rho$-$T$ curve fitting. This material is available free of charge via the Internet at http://pubs.acs.org.

AUTHOR INFORMATION

**Corresponding Author**

*E-mails: xujie@lanl.gov, hxu@lanl.gov, qxjia@lanl.gov

ACKNOWLEDGMENT

Xujie Lü acknowledges the J. Robert Oppenheimer Distinguished Fellowship supported by the Laboratory Directed Research and Development Program of Los Alamos National Laboratory. The work at Los Alamos National Laboratory was performed, in part, at the Center for Integrated Nanotechnologies, an Office of Science User Facility operated for the U.S. Department of Energy Office of Science, and in part by the U.S. Department of Energy, Office of Basic Energy Sciences, Division of Materials Sciences and Engineering. Sandia National Laboratories is a multi-program laboratory managed and operated by Sandia Corporation, a wholly owned subsidiary of Lockheed Martin Corporation, for the U.S. Department of Energy's National Nuclear Security Administration under contract DE-AC04-94AL85000. The authors thank Pamela Bowlan and Shan Guo for their helpful discussions.